\documentclass{pasa}
\usepackage{graphicx}
\usepackage{multirow}
\usepackage{svg}

\title{Performance of a Small Array of Imaging Air Cherenkov Telescopes sited in Australia}

\author[Simon Lee et al.]{Simon Lee$^1$, Sabrina Einecke$^1$, Gavin Rowell$^1$, Csaba Balazs$^2$, Jose A. Bellido$^1$, Shi Dai$^3$, Dominik Els\"asser$^4$, Miroslav Filipovi\'c$^3$, Violet M. Harvey$^1$, Padric McGee$^1$, Wolfgang Rhode$^4$, Steven Tingay$^5$, Martin White$^1$
\affil{$^1$School of Physical Sciences, University of Adelaide, Adelaide SA 5005, Australia}
\affil{$^2$School of Physics and Astronomy, Monash University, Melbourne VIC 3800, Australia}
\affil{$^3$School of Science, Western Sydney University, Locked Bag 1797, Penrith NSW 2751, Australia}
\affil{$^4$Department of Physics, TU Dortmund University, 44221 Dortmund, Germany}
\affil{$^5$International Centre for Radio Astronomy Research, Curtin University, GPO Box U1987, Perth, WA 6845, Australia}
}

\jid{PASA}
\doi{10.1017/pas.\the\year.xxx}
\jyear{\the\year}

\usepackage{aas_macros}
\usepackage{hyperref}
\hypersetup{colorlinks,citecolor=blue,linkcolor=blue,urlcolor=blue}

\usepackage{chngcntr}

\begin{document}
\begin{frontmatter}
\maketitle

\begin{abstract}
As TeV gamma-ray astronomy progresses into the era of the Cherenkov Telescope Array (CTA), there is a desire for the capacity to instantaneously follow up on transient phenomena and continuously monitor gamma-ray flux at energies above $10^{12}$\,eV.
To this end, a worldwide network of Imaging Air Cherenkov Telescopes (IACTs) is required to provide triggers for CTA observations and complementary continuous monitoring.
An IACT array sited in Australia would contribute significant coverage of the Southern Hemisphere sky.
Here, we investigate the suitability of a small IACT array and how different design factors influence its performance.
Monte Carlo simulations were produced based on the Small-Sized Telescope (SST) and Medium-Sized Telescope (MST) designs from CTA.
Angular resolution improved with larger baseline distances up to 277\,m between telescopes, and energy thresholds were lower at 1000\,m altitude than at 0\,m.
The $\sim$300\,GeV energy threshold of MSTs proved more suitable for observing transients than the $\sim$1.2\,TeV threshold of SSTs.
An array of four MSTs at 1000\,m was estimated to give a 5.7$\sigma$ detection of an RS Ophiuchi-like nova eruption from a 4-hour observation.
We conclude that an array of four MST-class IACTs at an Australian site would ideally complement the capabilities of CTA.
 
\end{abstract}

\end{frontmatter}

\section{INTRODUCTION }
\label{sec:intro}
Gamma-ray astronomy is a critical field for understanding the nature of extreme phenomena within and beyond our Galaxy.
However, in the very-high-energy (VHE) regime (10s of GeV to 100s of TeV) there is insufficient worldwide coverage to quickly follow up on or monitor transient and variable sources over a 24-hour period.

Imaging Air Cherenkov Telescopes (IACTs), such as \href{https://magic.mpp.mpg.de/}{MAGIC}, \href{https://www.mpi-hd.mpg.de/hfm/HESS/}{H.E.S.S}, \href{https://veritas.sao.arizona.edu/}{VERITAS}, and \href{https://www.isdc.unige.ch/fact/}{FACT}, measure the Cherenkov radiation from extensive air showers generated by gamma rays interacting with the Earth’s atmosphere.
These telescopes can detect VHE gamma rays with an angular resolution down to $\sim$0.05\textdegree.
They are very sensitive compared to alternate methods, allowing for measurements of flux variations on timescales as small as seconds.
The next-generation \href{https://www.cta-observatory.org/}{Cherenkov Telescope Array} (CTA) in its initial ``Alpha'' configuration will have 13 IACTs at its Northern Hemisphere site and 51 at its Southern Hemisphere site \citep{Sergijenko2021}.
These will provide dramatic improvements to sensitivity across the VHE regime \citep{2018CTA}.
The limitations of IACTs are their comparatively narrow field-of-view and their optical detection method, which restricts observations to night time.

Water Cherenkov detectors (WCDs), such as those used in \href{https://www.hawc-observatory.org/}{HAWC}, \href{https://english.ihep.cas.cn/lhaaso/}{LHAASO}, and the upcoming \href{https://www.swgo.org}{SWGO}, detect Cherenkov light from charged particles passing through large bodies of water instead of air.
The main benefits of this method are the very wide field-of-view achievable and the ability to run 24 hours a day, allowing many sources to be monitored simultaneously for long time periods.
Compared to IACTs, WCDs are orders of magnitude less sensitive for a given observation time, and their angular resolution quickly deteriorates below 10 TeV \citep{Wang2018}.
This makes them less capable at detecting faint transients, reconstructing spectra for short-lived events, and monitoring flux variations on the scale of hours, minutes, or seconds.

The \href{https://glast.sites.stanford.edu/}{Large Area Telescope} (LAT) on the \emph{Fermi} satellite directly detects gamma rays with a collection area of $\sim$1\,m$^2$.
It has a wide field-of-view and can observe the whole sky multiple times per day as it orbits the Earth \citep{Atwood2013}.
The Fermi All-sky Variability Analysis (FAVA) Catalog \citep{Abdollahi2017} so far comprises thousands of recorded gamma-ray flares from hundreds of associated transient sources, including concurrent gamma-ray detections with multi-messenger transient events such as the gravitational waves from GW170817 \citep{Ajello2018} and the high-energy neutrinos from TXS0506+056 \citep{IceCube2018}.
\emph{Fermi}-LAT is however limited to sub-TeV energies, with an energy range between 20\,MeV and 300\,GeV.
Because of its small effective area, flux variations are typically only measured on a daily or weekly basis, and spectral reconstructions on these timescales are usually not possible.

\begin{figure}[t]
    \begin{center}
        \includegraphics[width=1.0\columnwidth]{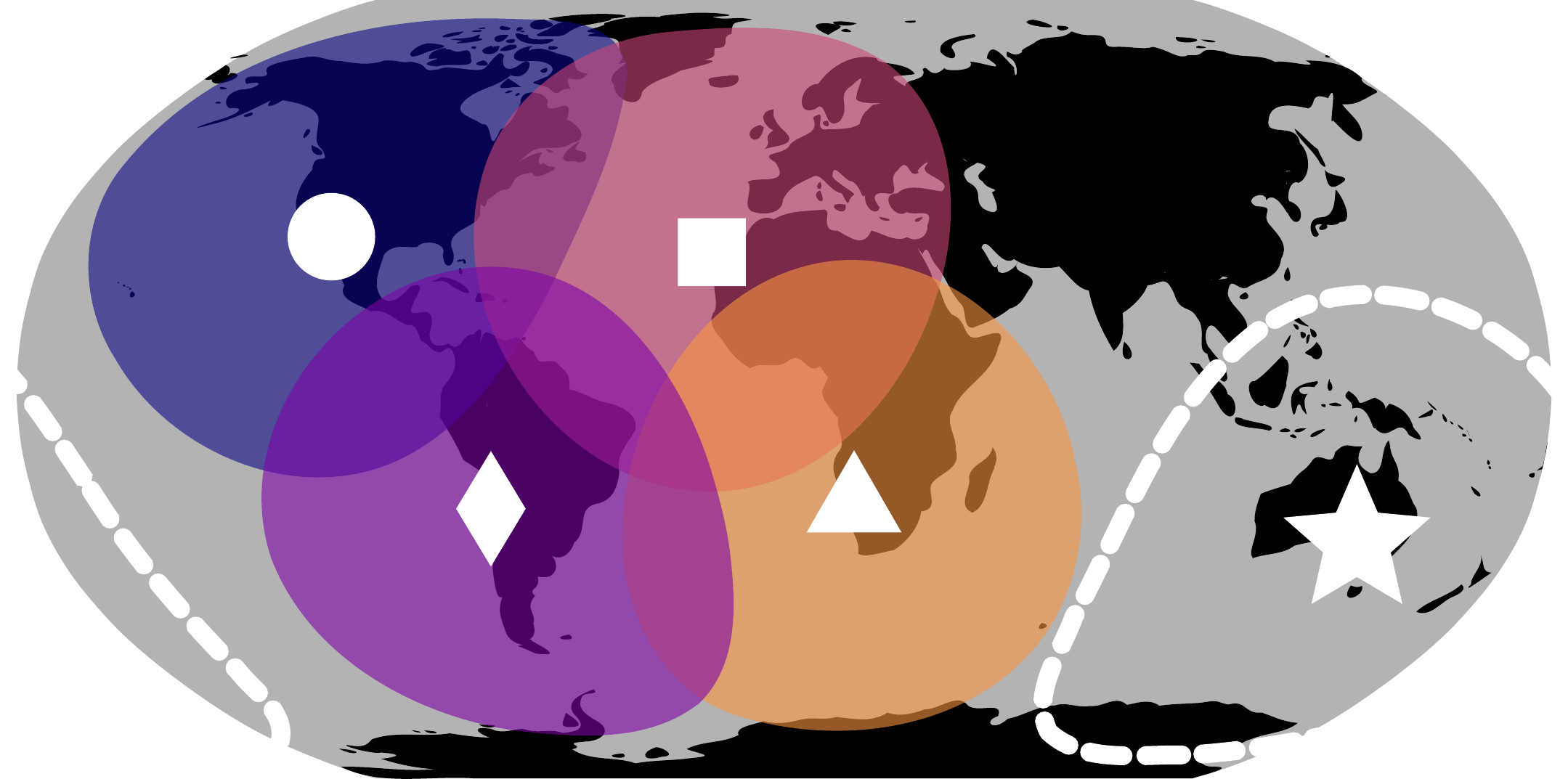}
        \caption{Sky coverage $<$50\textdegree \ zenith of the IACTs VERITAS ({\Large $\bullet$}), MAGIC/FACT/CTA-North ($\blacksquare$), CTA-South ({$\blacklozenge$}), and H.E.S.S. ({\large $\blacktriangle$}).
        The necessity for an Australian site ({\large $\bigstar$}) for obtaining 24-hour sky coverage is readily apparent.}
        \label{world}
    \end{center}
\end{figure}

To more deeply investigate the behaviours of transient phenomena and variable sources, a worldwide network of IACTs would be necessary.
This idea has been proposed several times over the past decades.
\citet{Lorenz2005} suggested a Northern Hemisphere network of pre-existing small Cherenkov telescopes to monitor the sky for active galactic nuclei (AGN) flares, as newer and larger telescopes were focused on other studies.
The Dedicated Worldwide AGN Research Facility (DWARF) \citep{2009icrc.conf.1452B} sought to achieve full-sky coverage with existing and new telescopes, including re-purposing equipment from HEGRA.
Most recently the Cherenkov Telescope Ring (CTR) \citep{Ruhe2019} was presented as an idea to establish new telescope sites to complement CTA into the future.
The performance of small IACT arrays for TeV observations has also been previously studied \citep{Plyasheshnikov2000,Kifune2001,yoshikoshi2005,Colin2007,Stamatescu2011}, including at a low-altitude site in Australia \citep{Rowell2008}.
\\

\begin{figure}[t]
    \begin{center}
        \includegraphics[width=1.0\columnwidth]{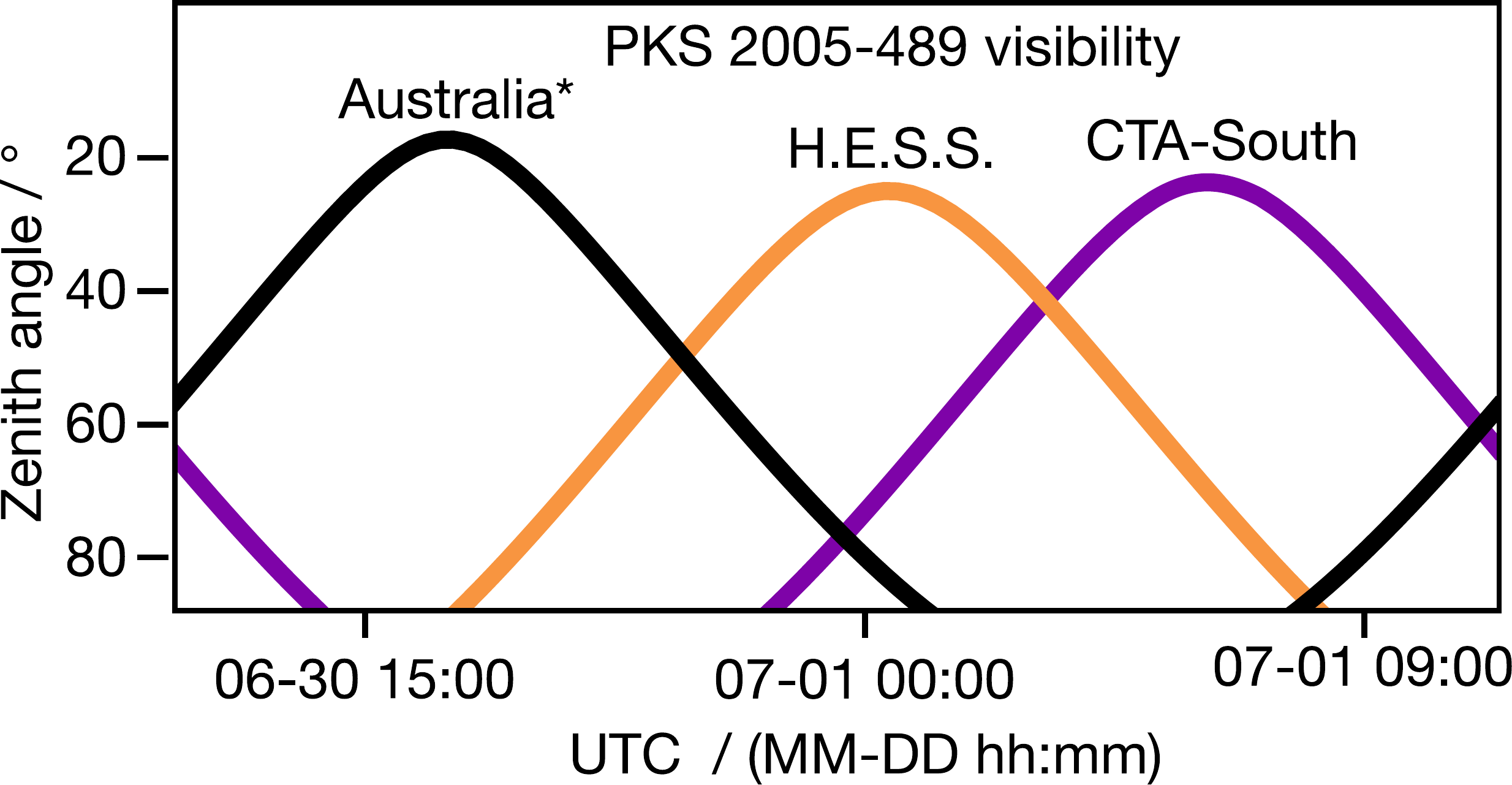}
        \caption{Visibility of the blazar PKS 2005-489 by different IACT sites over 24 hours.
        A site in Australia (*Arkaroola: 30.3\textdegree\,S, 139.3\textdegree\,E) would complement H.E.S.S. and the upcoming CTA-South site for continuous Southern hemisphere coverage.}
        \label{visibility}
    \end{center}
\end{figure}

The currently running IACT arrays include MAGIC and FACT in the Canary Islands (Spain), H.E.S.S. in Namibia, and VERITAS in the United States.
Even with the upcoming CTA-South in Chile and CTA-North in the Canary Islands, this does not provide continuous full-sky coverage (see \autoref{world}).
An IACT array in Australia would help fill in the gap and allow for follow-up observations of the southern sky at any time (see \autoref{visibility}) continuing in the legacy of
previous Australia-sited IACTs such as BIGRAT \citep{Clay1989}, The University of Durham telescopes \citep{Armstrong1999}, and CANGAROO \citep{Enomoto2002}.
With the aim of expanding the science capabilities of CTA, such an array could provide triggers for further observation, and continue observing transient events triggered by CTA over the following day.
Furthermore it could contribute triggers to Australian optical and radio observatories, particularly the Australian component of the upcoming Square Kilometre Array (SKA) next-generation radio telescope \citep{Johnston2007}.

\begin{table*}[t]
\caption{Simulation parameters used for each site altitude (0\,m \& 1000\,m).
Particle energies were drawn following the relationship $dN/dE \propto E^\Gamma$.
Diffuse emission was generated within a ``view cone'' of radius $R_{\mathrm{view\,cone}}$.
The shower core positions were evenly distributed in a circular area of radius $R_{\mathrm{scatter}}$.
Every shower was re-used with their core positions varied, providing alternate views of the shower (Shower re-use).}
\label{settings}
\centering
\begin{tabular}{lcccccc}
\hline
\begin{tabular}[c]{@{}l@{}}Particle  type\end{tabular} & \begin{tabular}[c]{@{}l@{}}Energy range\end{tabular} & \begin{tabular}[c]{@{}l@{}}$\Gamma$\end{tabular} & \begin{tabular}[c]{@{}l@{}}Shower re-use\end{tabular} & \begin{tabular}[c]{@{}l@{}}$R_{\mathrm{view\,cone}}$\end{tabular} & \begin{tabular}[c]{@{}l@{}}$R_{\mathrm{scatter}}$\end{tabular} & \begin{tabular}[c]{@{}l@{}}Total events\end{tabular} \\ \hline
\begin{tabular}[c]{@{}l@{}}Gamma (on-axis)\end{tabular} & \begin{tabular}[c]{@{}l@{}}0.06 - 300\,TeV\end{tabular} & -1.3 & 20 & 0\,deg & 1060\,m & 1.3 million \\
\begin{tabular}[c]{@{}l@{}}Gamma (diffuse)\end{tabular} & \begin{tabular}[c]{@{}l@{}}0.06 - 300\,TeV\end{tabular} & -1.3 & 20 & 10\,deg & 1560\,m & 20 million \\
Proton (diffuse) & \begin{tabular}[c]{@{}l@{}}0.06 - 300\,TeV\end{tabular} & -1.3 & 20 & 10\,deg & 1560\,m & 126 million \\ \hline
\end{tabular}
\end{table*}

Such a project could have a wide range of scientific objectives.
For example, AGN make up the largest fraction of known TeV sources.
Blazars in particular (AGN whose jets point towards Earth) give opportunities to investigate the particle acceleration processes, inner-jet dynamics, and the central engines of AGN \citep{Sikora1997, Filipovic2021}.
The continuous monitoring capabilities provided by such a network would allow further study into particular AGN of interest, generating time-dependent spectral energy distributions, densely sampled unbiased light curves, and increasing the chances of detecting AGN flares (and triggering other telescopes for multi-wavelength observations or for follow-ups by more sensitive gamma-ray telescopes).
The increased observation time could additionally be used to help demystify the large collection of as-yet unclassified hard-spectrum gamma-ray sources in the FAVA catalogue, many of which might extend into TeV energies \citep{Abdollahi2017}.

Gamma-ray bursts (GRBs) are still poorly understood phenomena, and recent detections have shown them capable of reaching TeV energies \citep{MAGIC2019,HESS2021}.
Obtaining time-resolved spectra of GRBs is necessary to determine their underlying emission mechanisms.
A worldwide IACT network would allow for quickly responding to GRB triggers from other telescopes, and give higher chances of detecting GRBs due to increased sky monitoring time.
This would give more opportunities to investigate their mechanisms through observing gamma-ray flux variation in their immediate aftermath and correlating this with other wavelengths.
Gravitational wave astronomy continues to progress as a significant field of discovery, and the 24-hour availability for instantaneous gamma-ray follow-up would be instrumental to studying the nature of neutron star mergers \citep{Ajello2018}.
A recent outburst of the nova RS Ophiuchi observed by H.E.S.S. and MAGIC has shown that novae can also be sources of $>$100\,GeV gamma rays \citep{HESS2022, MAGIC2022}, providing exciting opportunities as an emerging VHE source class, which the proposed network could contribute to with its capacity for instantaneous follow-up.

The design of a new IACT array in Australia should be informed by what performance is achievable.
In particular, the availability of high-altitude sites suitable for optical-based telescopes in Australia is limited.
Siding Spring Observatory for instance (a hub for many optical telescopes) has a maximum altitude of 1165\,m, whereas both CTA sites will be situated above 2000\,m.
Here, we therefore study the performance of such an array dependent on altitude, geometric layout, number of telescopes, and telescope designs based on those from CTA.

\section{METHODS}

In this study we investigated the performance achievable by a small array of up to four IACTs at altitudes of 0\,m and 1000\,m.
The primary performance metric considered was an array's differential sensitivity to a steady, point-like gamma-ray source.
Monte Carlo simulations were produced to replicate such observations from which this and other performance metrics were derived.

\subsection{Simulations}

Gamma rays and charged cosmic rays (protons and heavier nuclei) generate extended air showers from colliding with atmospheric molecules, resulting in a cascade of particles showering towards the ground and generating Cherenkov radiation.
This light is detected by IACTs, with cosmic ray showers occurring at least $\sim$1000 times more frequently than those from gamma rays.
For this study, Monte Carlo simulations of gamma-ray showers, protons showers, and the Cherenkov light they produce were made with \verb|CORSIKA|\footnote{Version 7.7100 with the QGSJET II-03 interaction model.} \citep{1998cmcc}.

Gamma rays were simulated originating from a point north of the array, and at a zenith angle of 20\textdegree\ where optimal sensitivity could be expected.
The telescopes were aimed with a 1\textdegree\ offset from this point.
Diffuse emission of protons (for background) and further gamma rays (for unbiased reconstruction models) were also simulated.
The geomagnetic field of a site located in Arkaroola (30.3\textdegree\,S, 139.3\textdegree\,E) was used to emulate a potential Australian site, generated by the \verb|Geomag 7.0| software \citep{Alken2021}.
\autoref{settings} presents the simulation settings used for both site altitudes of 0\,m and 1000\,m.

\begin{figure}[t]
    \begin{center}
        \includegraphics[width=0.7\columnwidth]{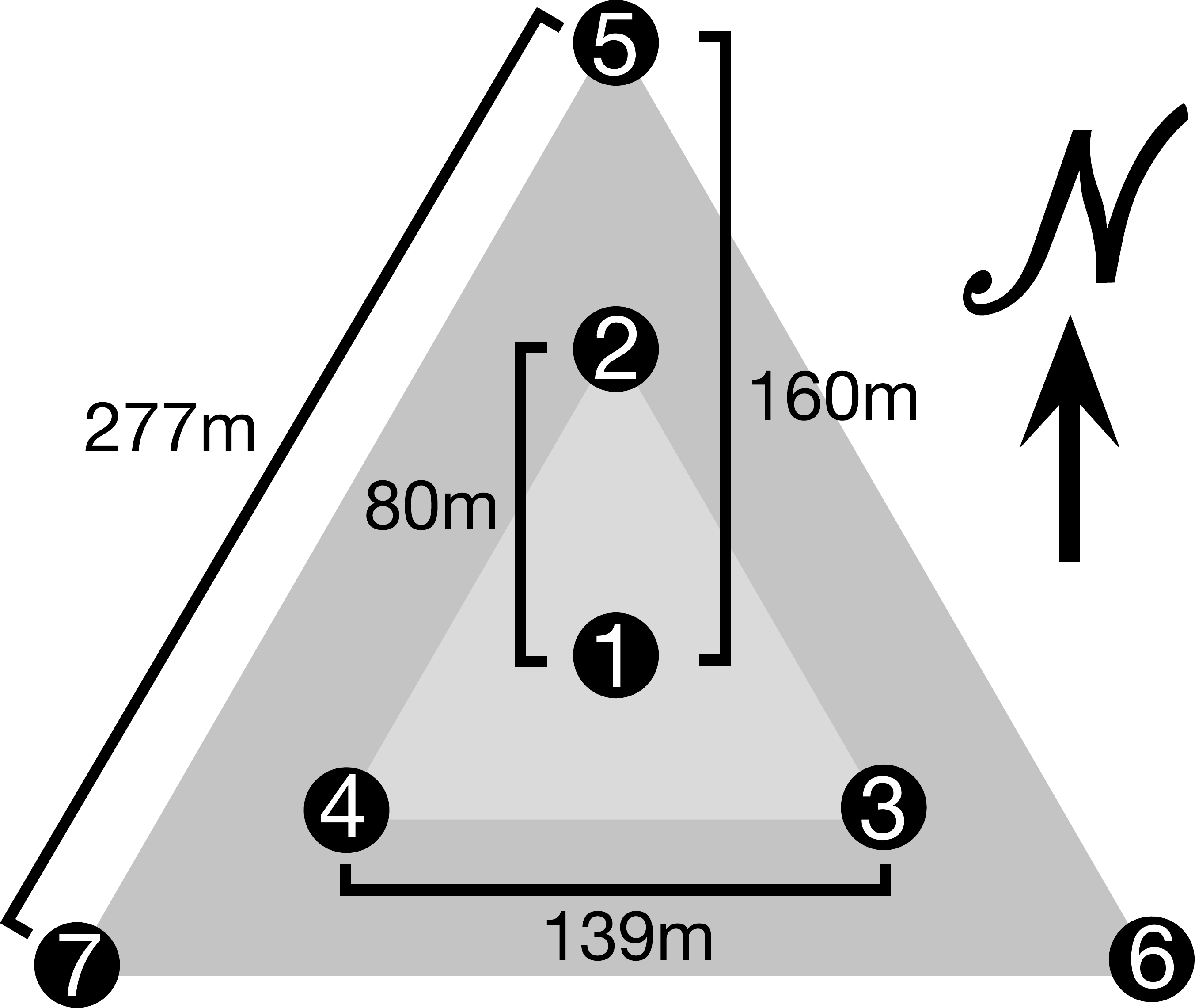}
        \caption{Arrangement of IACTs (shown as numbers) used in simulations, allowing for multiple different configurations of baseline distances and number of telescopes to be studied.}
        \label{tels}
    \end{center}
\end{figure}

The Cherenkov photons generated by the air showers were passed into the \verb|sim_telarray| \citep{Bernloehr2008} package to simulate IACTs observing the showers.
For each photomultiplier converting photons to electrons in a camera sensor, this produces a waveform within a short ($\sim$100\,ns) time window around a triggered event.
The package takes into account aspects such as mirror reflectivity, telescope structure shadowing the mirrors, night sky background, trigger conditions for event recording, and the quantum efficiency of the camera sensor.
The telescopes were arranged with a central telescope surrounded by three more in an equilateral triangle each 80\,m away, and another three 160\,m away (see \autoref{tels}).
This allowed for the study of setups from one to four telescopes with a variety of baseline distances (the maximum distance between telescopes in an array).
As a basis for testing, the state-of-the-art CTA Prod-5\footnote{Version 2020-06-28.} \citep{CTAO2021} designs of the 12-metre Medium-Sized Telescope (MST) and 4-metre Small-Sized Telescope (SST) were chosen to be simulated as affordable solutions to provide good sensitivity above 0.1\,TeV.

The altitude at which an IACT operates will affect various aspects of its observations \citep{Hassan2017}.
Most of the Cherenkov light in an extensive air shower is produced 5 -- 10\,km above ground level.
As it propagates towards the ground, the Cherenkov light pool spreads out, covering more area but with lower photon density.
For showers whose core lands close to a telescope, higher altitude sites will produce images with greater intensity, generally leading to better event reconstruction and a lower energy threshold.
However, the smaller Cherenkov light pool at higher altitudes results in a lower likelihood of showers being detected by multiple telescopes, and more distant showers would be seen with lower photon density, or not at all. 
Simulation runs were thus made at two different heights above sea level (0\,m and 1000\,m) to study the variation in performance given the site altitudes available in Australia.

\subsection{Analysis}

Tools from \verb|ctapipe| \citep{Kosack2021}, a prototype data processing framework for CTA, were used to perform the low-level event processing.
Pixel intensities and arrival times were extracted from their waveforms using the \verb|Neighbour Peak Window Sum| method, which chooses extraction windows dependent on the surrounding pixel waveforms.
The extraction windows were optimised for accurate noise extraction\footnote{Width/shift values of 6/3 samples were used for SSTs, and 4/1 samples for MSTs.}.
Images were cleaned to remove pixels without Cherenkov light using a combination of the two-level tail cut approach\footnote{Core/boundary thresholds were chosen at 10/5 photoelectrons for MSTs and 3/1.5 for SSTs.} (to remove dim pixels not adjacent to bright pixels) and \verb|time delta cleaning| (to remove pixels with arrival times not coincident with those adjacent).
Cleaned images were parameterised by the second-moment Hillas analysis \citep{hillas1985} to be used for removing low-quality images, energy reconstruction, direction reconstruction, and gamma/hadron separation.

The following quality cuts were required to remove low-quality images:
\begin{itemize}
    \item Leakage\footnote{An analogue for Cherenkov ellipse truncation by the edge of the camera sensor, defined as the ratio (post-cleaning) between summed pixel intensities at the edge of the camera and the total summed intensity.}: < 0.2
    \item Total photoelectrons: > 70 for MSTs, 30 for SSTs
    \item Surviving pixels: > 5
    \item Number of islands\footnote{Disjoint clusters of pixels post-cleaning.}: < 4
\end{itemize}

Training and applying models for event reconstruction were performed with \verb|aict-tools| \citep{aict-tools}.
Models for reconstructing energy, direction, and event conformity to a gamma ray (gamma score) were produced using random forests (RFs), a well-established technique for event reconstruction with IACTs \citep{Albert2008}.
All available diffuse gamma rays were used for creating the models, and an equal number of diffuse protons were separated from the dataset to train on.
The models were applied to the point-source gamma rays and the remaining diffuse protons.
Events detected by multiple telescopes were reconstructed from the mean of individual telescope reconstructions, weighted by image intensity (total number of extracted photoelectrons).

Additionally, a geometric direction reconstruction was performed along the weighted intersection of planes passing through the major axes of cleaned images, resulting in both geometric and RF-reconstructed directions and $\theta$ values (the distance between true and reconstructed source position).
For a given array setup, there was a geometrically-reconstructed impact distance (distance from the array centre to the shower core on the ground) beyond which the geometric direction reconstruction performed worse on average than the RF direction reconstruction due to the more acute stereo angle (the angle formed by the projection of the shower axis in two cameras).
Thus, for each telescope arrangement, geometric direction reconstruction was used within this calculated distance, and RF direction reconstruction was used beyond it (values presented in \autoref{rfcuts}).

\begin{figure*}[t]
    \begin{center}
      \centering
          \includegraphics[width=0.8\textwidth]{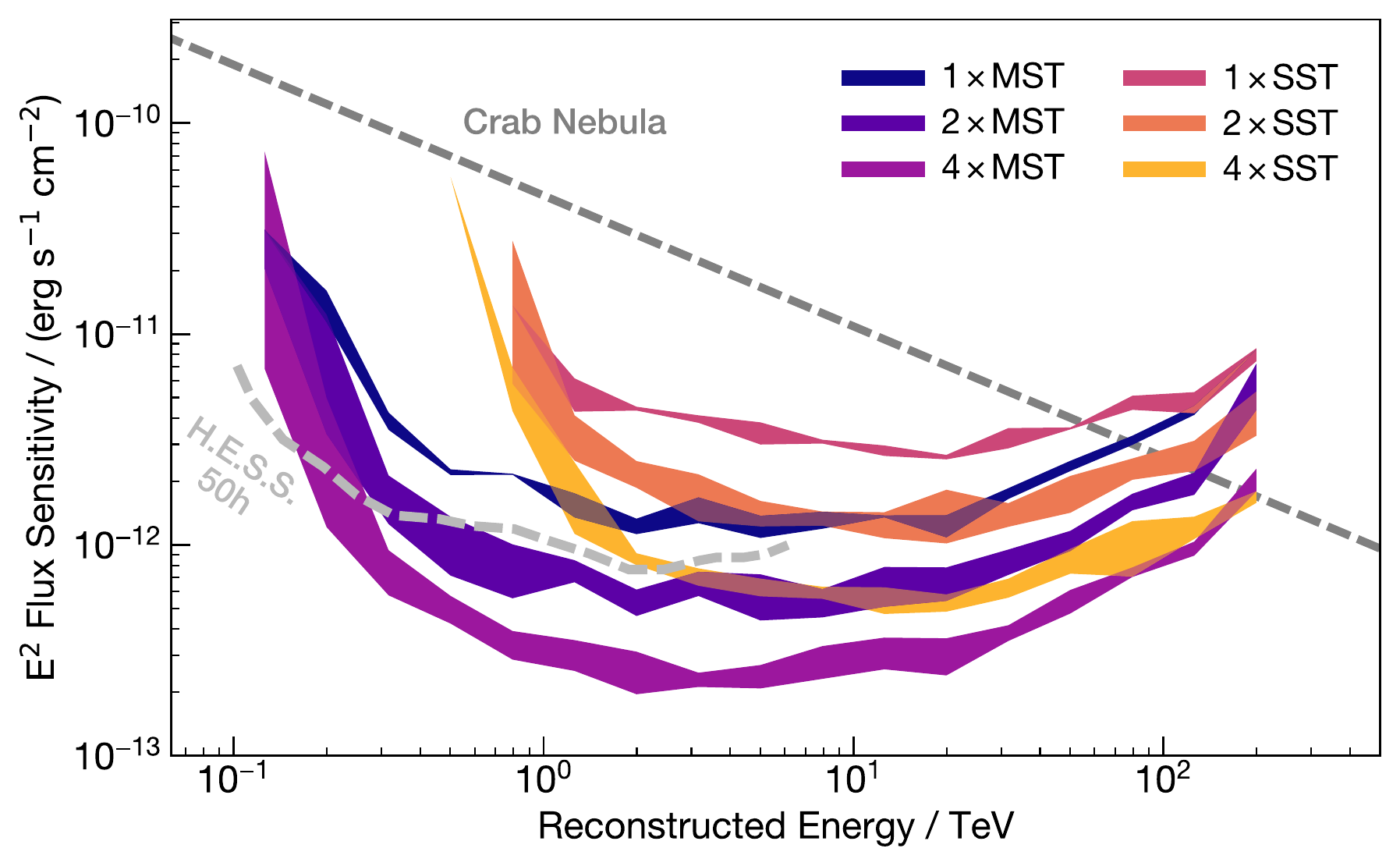}
          \caption{50-hour differential point-source flux sensitivity for a 5$\sigma$ detection as a function of reconstructed gamma-ray energy.
          Bands represent the range of sensitivities across the studied altitudes (0\,m and 1000\,m) and baseline distances (80\,m to 277\,m).
          Cuts on gamma score and $\theta^{2}$ were applied for each energy bin to optimise sensitivity for each array setup.
          No cuts on the number of telescopes triggered were applied.
          The H.E.S.S. 50-hour sensitivity curve is shown for comparison \citep{holler2015}.}
          \label{sensitivity}
    \end{center}
\end{figure*}

A telescope array's differential sensitivity for a given energy range is defined as the minimum flux required for a point source to be observed with a significance of 5\,$\sigma$ after 50 hours.
The significance was found using the Li \& Ma method \citep{Li1983} with one on-region and five off-regions (chosen equidistant around the camera centre).
At least 10 excess on-region counts, 10 counts in the off regions, and an excess to background ratio of $>\frac{1}{20}$ was required for each energy bin, as per the standards adopted by CTA \citep{Hassan2017} and others.
For each array setup, a minimum gamma score and maximum $\theta^{2}$ (where $\theta$ is the radius of on- and off-regions) were chosen to optimise sensitivity for each energy bin.
The effective area was determined for each energy bin by multiplying the area over which the showers were thrown at the array by the ratio of the number of observed showers (post-cuts) and the total number of thrown showers.
The angular resolution was calculated as the 68\textsuperscript{th} percentile of $\theta$ values (post-cuts) for each energy bin.
As differential point-source sensitivity was the primary performance metric used to compare setups, the gamma score and $\theta^{2}$ cuts found to optimise sensitivity for each energy bin were applied to the datasets used in both angular resolution and effective area plots.
Depending on the desired performance criteria, cuts could instead be chosen to optimise for angular resolution (with stronger gamma score cuts) or for effective area (with the loosening of either gamma score or $\theta^{2}$ cuts).

\section{ARRAY PERFORMANCE}
\begin{figure}[t]
\begin{center}
  \centering
  \includegraphics[width=0.9\columnwidth]{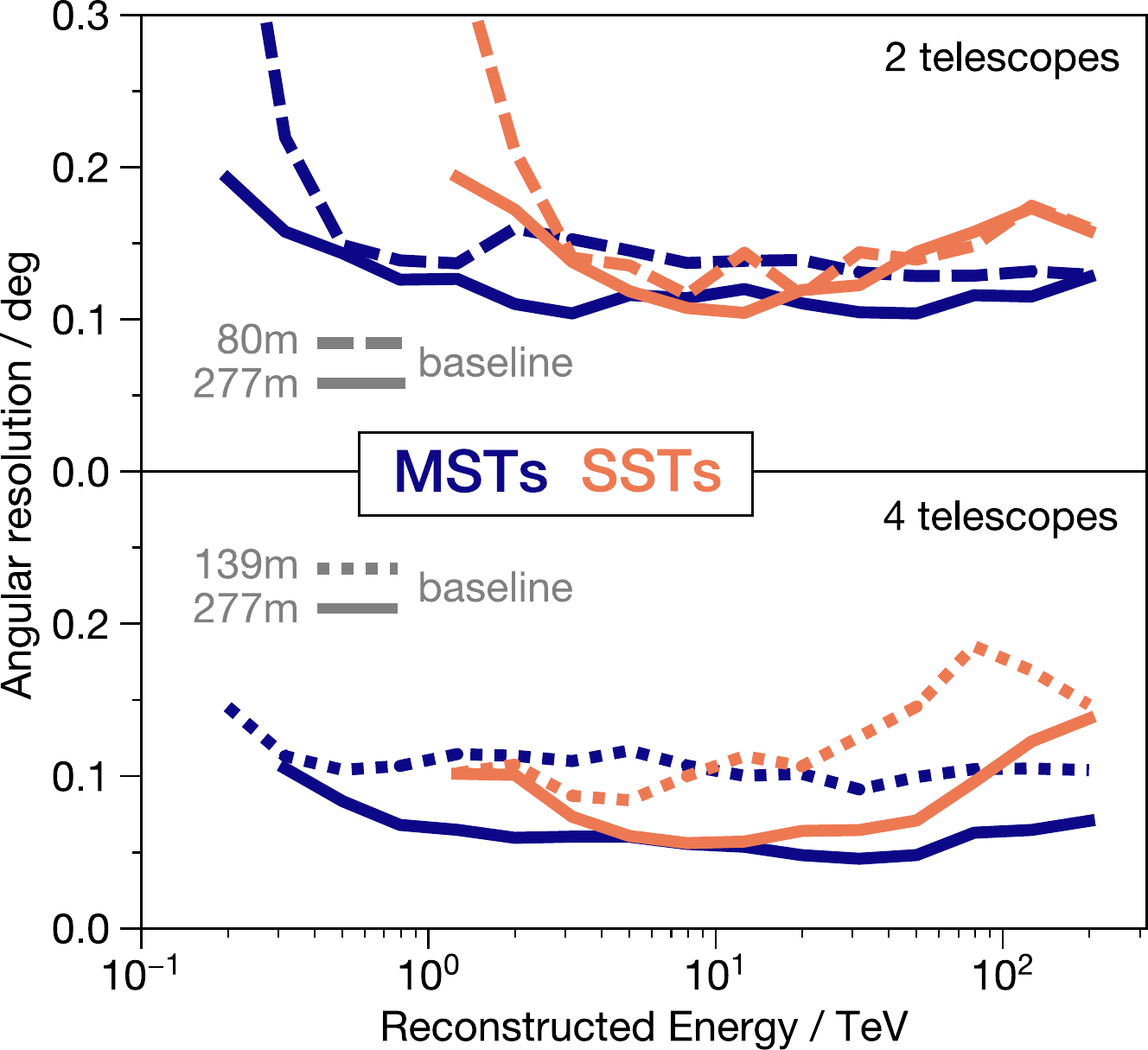}
  \caption{Angular resolution as a function of reconstructed gamma-ray energy for 0\,m altitude arrays.
  Gamma score cuts optimised for sensitivity per energy bin were applied.
  Events were chosen where all telescopes triggered, otherwise monoscopic events dominated and results were similar to a single-telescope setup.
  The corresponding sensitivity was very similar between equivalent arrays of different baselines (see \autoref{sens_baseline}).
  }\label{baseline}
\end{center}
\end{figure}

\begin{figure}[t]
    \begin{center}
      \centering
          \includegraphics[width=0.9\columnwidth]{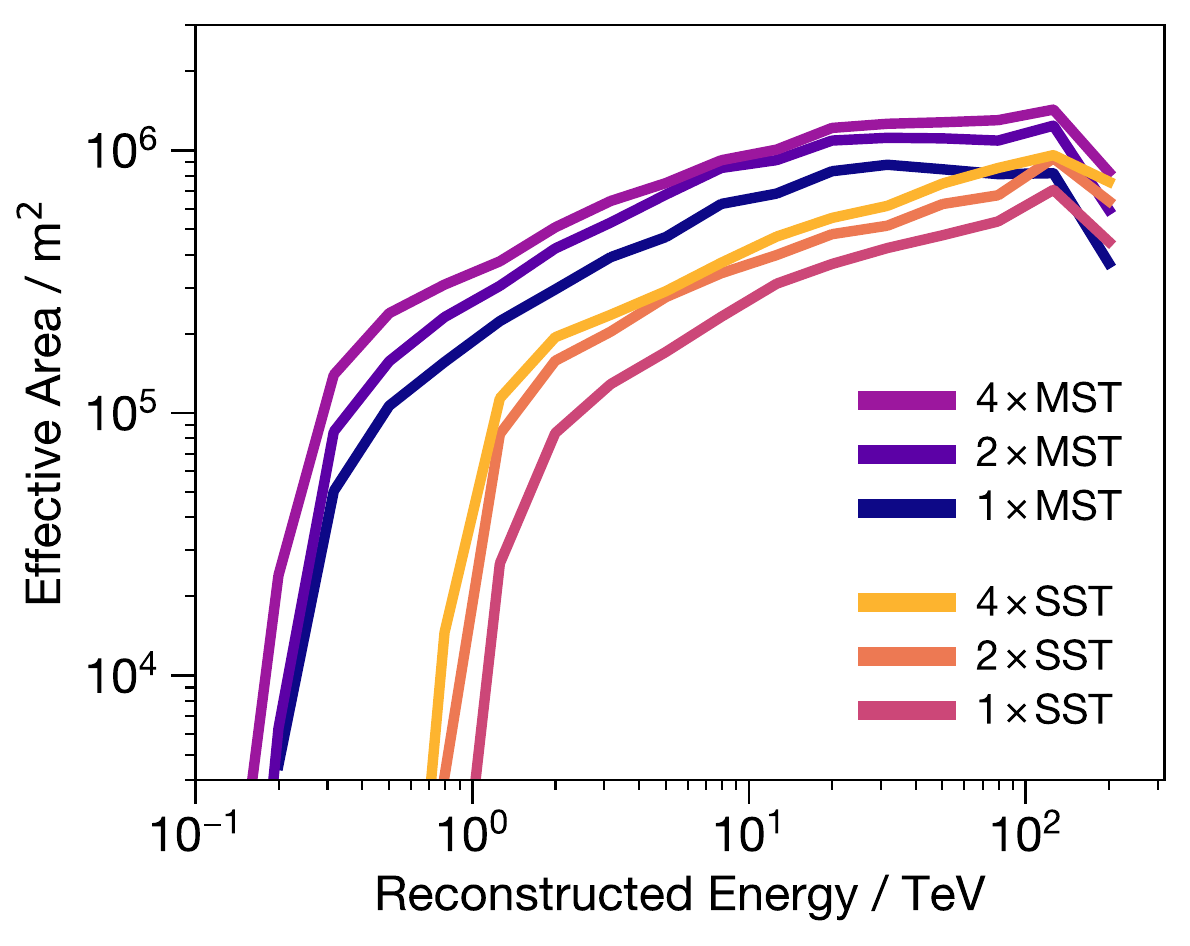}
          \caption{Effective area as a function of reconstructed gamma-ray energy for arrays at 0\,m altitude with a 277\,m baseline.
          $\theta^{2}$ and gamma score cuts optimised for sensitivity per energy bin were applied, with no cuts on the number of telescopes triggered.}
          \label{effarea}
    \end{center}
\end{figure}

Performance comparisons are shown in Figures \ref{sensitivity} through \ref{sensvtime}.
Due to the large number of setups compared in this study, the specific effects of altitude, baseline, number of telescopes, and telescope size are discussed, with the range of performance variation presented as bands in \autoref{sensitivity}.

\subsection{Number of Telescopes per Array}

The number of telescopes in an array was varied from one to four.
As expected there were improvements to all performance metrics with more telescopes across all altitudes, baseline distances, and telescope sizes.
Increasing from 1 to 2 and 2 to 4 telescopes provided an approximately 2.5 times improvement in sensitivity (see \autoref{sensitivity}), roughly 0.05\textdegree\ better angular resolution (see \autoref{baseline}), and $\sim$30\% larger effective areas (see \autoref{effarea}).
This is to be expected as more telescopes allow for larger ground coverage, more accurate stereoscopic direction reconstruction, and more estimates of particle type, source position, and energy for producing weighted-average predictions.
 
\subsection{Altitude}

Arrays at 1000\,m altitude had small improvements in low-energy performance over those at 0\,m (see \autoref{sens_alt}).
The lowest energy bin in the sensitivity curves of all MST arrays extended to $\sim$120\,GeV when at 1000\,m altitude.
Compared to those at 0\,m, arrays with four MSTs had an order of magnitude improvement at this energy, and the lowest energy bin for arrays with one or two MSTs was at $\sim$200\,GeV.
SST arrays had similar results, such as with an additional lower energy bin down to $\sim$930\,GeV for one SST, and down to $\sim$580\,GeV for four SSTs with a 138\,m baseline.
Angular resolution at low energies improved by up to 50\% for both MST and SST arrays with two telescopes 80\,m apart, but otherwise there was negligible performance variation.
Conversely, all arrays at 0\,m altitude had up to 25\% larger effective area above 1\,TeV and a higher energy bin in the sensitivity of monoscopic MST setups (extending up to $\sim$230\,TeV).
This can be understood by the fact that Cherenkov light pools become broader and less photon-dense as they propagate.

\subsection{Array Baseline}

The performance with respect to baseline distance was compared, varying between 80\,m (only for two-telescope arrays), 139\,m, and 277\,m.
When increasing the baseline from 80\,m to 277\,m, arrays of two telescopes had improvements in angular resolution of $\sim$50\% near the energy threshold (down to below 0.2\textdegree), and MSTs showed up to 25\% improvement above 1\,TeV (down to almost 0.1\textdegree) (see \autoref{baseline}).
3- and 4-telescope arrays had up to two-fold improvements in angular resolution across most energies (down to $\sim$0.05\textdegree) when doubling the triangular outer baseline from 139\,m to 277\,m.
These results were due to the more perpendicular stereo angle allowing for improved geometric direction reconstruction.
Larger baselines also corresponded to increases in effective area in all arrays across all energies (by $\sim$20-40\%).

The effect of baseline distance on sensitivity was most notable in 4-telescope MST arrays at 1000\,m altitude, with improvements of up to 50\% below 700\,GeV with a 139\,m baseline compared to one of 277\,m, and a worsening of performance between 700\,GeV and 5\,TeV for the same comparison (see \autoref{sens_baseline}).
Differences in sensitivity performance with respect to baseline for other arrays were otherwise small.
In the 1000\,m SST arrays, shorter baselines also resulted in improved angular resolution below $\sim$3\,TeV.
For these comparatively dim showers near the energy threshold, this result can be understood as an effect of smaller Cherenkov light pools at higher altitudes.
With a larger baseline, a higher proportion of events are those that land between the telescopes and have obtuse stereo angles.
This results in poorer geometric reconstruction on average, and up to 50\% worse angular resolution at these lower energies.

\subsection{SST vs MST}

The largest difference in performance between telescope types was in energy threshold.
SST arrays had energy thresholds of $\sim$1.2\,TeV whereas MST arrays had thresholds of $\sim$300\,GeV.
Angular resolution was comparable for a given number of telescopes, with marginal improvements for stereoscopic MST arrays over stereoscopic SST arrays (see \autoref{baseline}).
A single MST provided a larger effective area below 40\,TeV than four SSTs, and for a given number of telescopes an equivalent MST array improved on it four-fold (see \autoref{effarea}).
Below 10\,TeV one MST had similar sensitivity to two SSTs and two MSTs were comparable to four SSTs (see \autoref{sensitivity}).

\subsection{Sensitivity vs Time}

\begin{figure}[t]
    \begin{center}
          \centering
          \includegraphics[width=0.9\columnwidth]{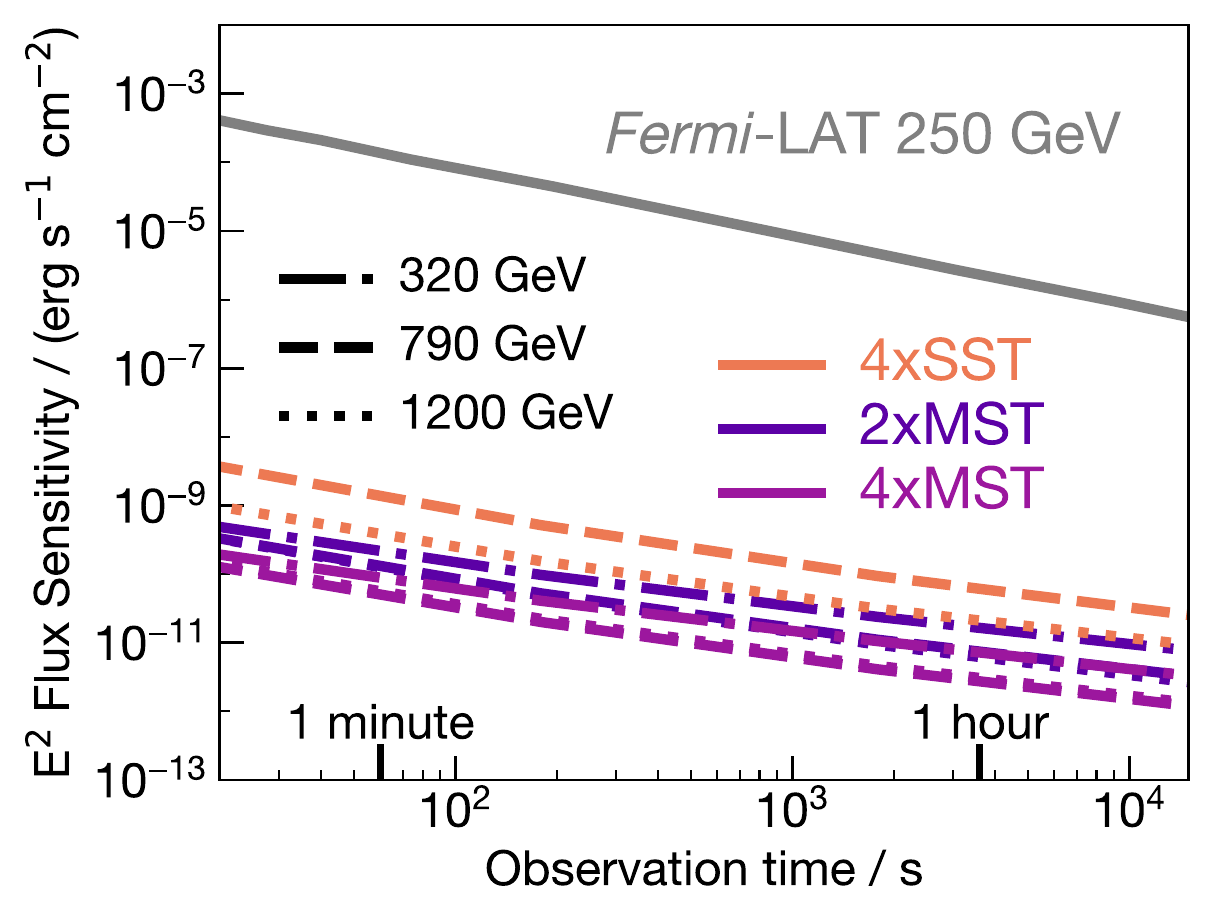}
          \caption{Differential point-source flux sensitivity for a 5$\sigma$ detection as a function of observation time for selected energy bins for arrays at 0\,m altitude with baselines of 277\,m.
          Cuts on gamma score and $\theta^{2}$ were applied for each energy bin to optimise sensitivity for each array setup.
          The SST lacks a 320\,GeV line as it is outside the detectable energy range.
          The sensitivity of \emph{Fermi}-LAT (grey) is shown for comparison.}
          \label{sensvtime}
    \end{center}
\end{figure}

\autoref{sensvtime} shows the lowest flux detectable at 5$\sigma$ significance as a function of time.
This metric is of note as it pertains to the ability to probe short-timescale flux variations and transient events.
MST arrays were several orders of magnitude more sensitive than \emph{Fermi}-LAT at $\sim$320\,GeV.
The SST array’s sensitivity does not extend this low, highlighting the main benefit of the MST array being its lower energy threshold.
WCDs such as those to be employed at SWGO vary across sensitivity between $\sim$10\textsuperscript{2}--10\textsuperscript{-3} erg s\textsuperscript{-1} cm\textsuperscript{-2} at $\sim$300\,GeV in the temporal range shown  \citep{Albert2019}.
This plot clearly highlights the benefits of an IACT array over alternate methods for observing faint, short-lived, or quickly varying transient phenomena.

\section{DISCUSSION}
\begin{figure}[t]
    \begin{center}
          \centering
          \includegraphics[width=0.95\columnwidth]{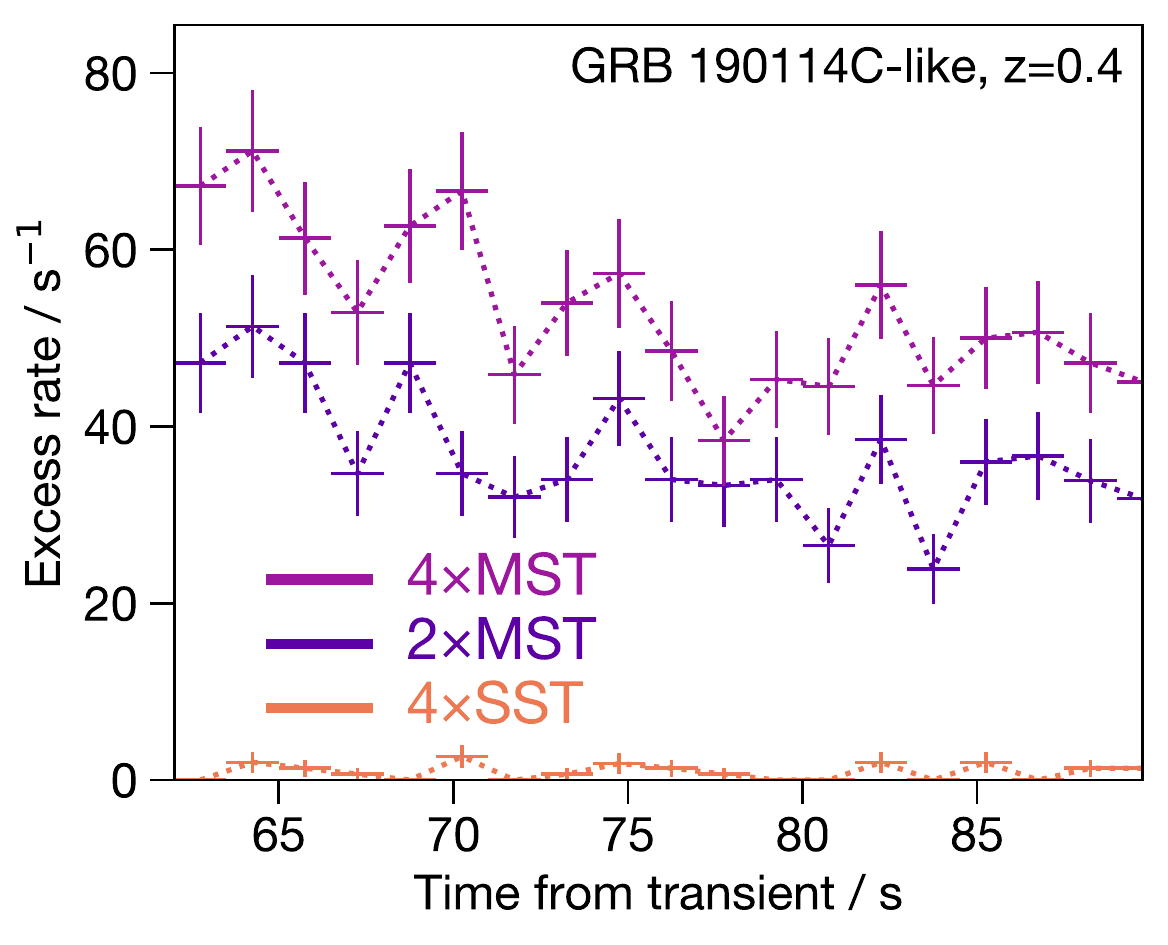}
          \caption{Estimated light curves for a GRB 190114C-like event for arrays at 0\,m altitude with baselines of 277\,m.
          The vertical bars show standard deviation, and horizontal bars show observation time per bin.
          Due to the 1.5-second binning, the mean background rate for all arrays was 0 protons and electrons per second.}
          \label{lightcurve}
    \end{center}
\end{figure}

\begin{figure}[ht]
    \begin{center}
          \centering
          \includegraphics[width=0.9\columnwidth]{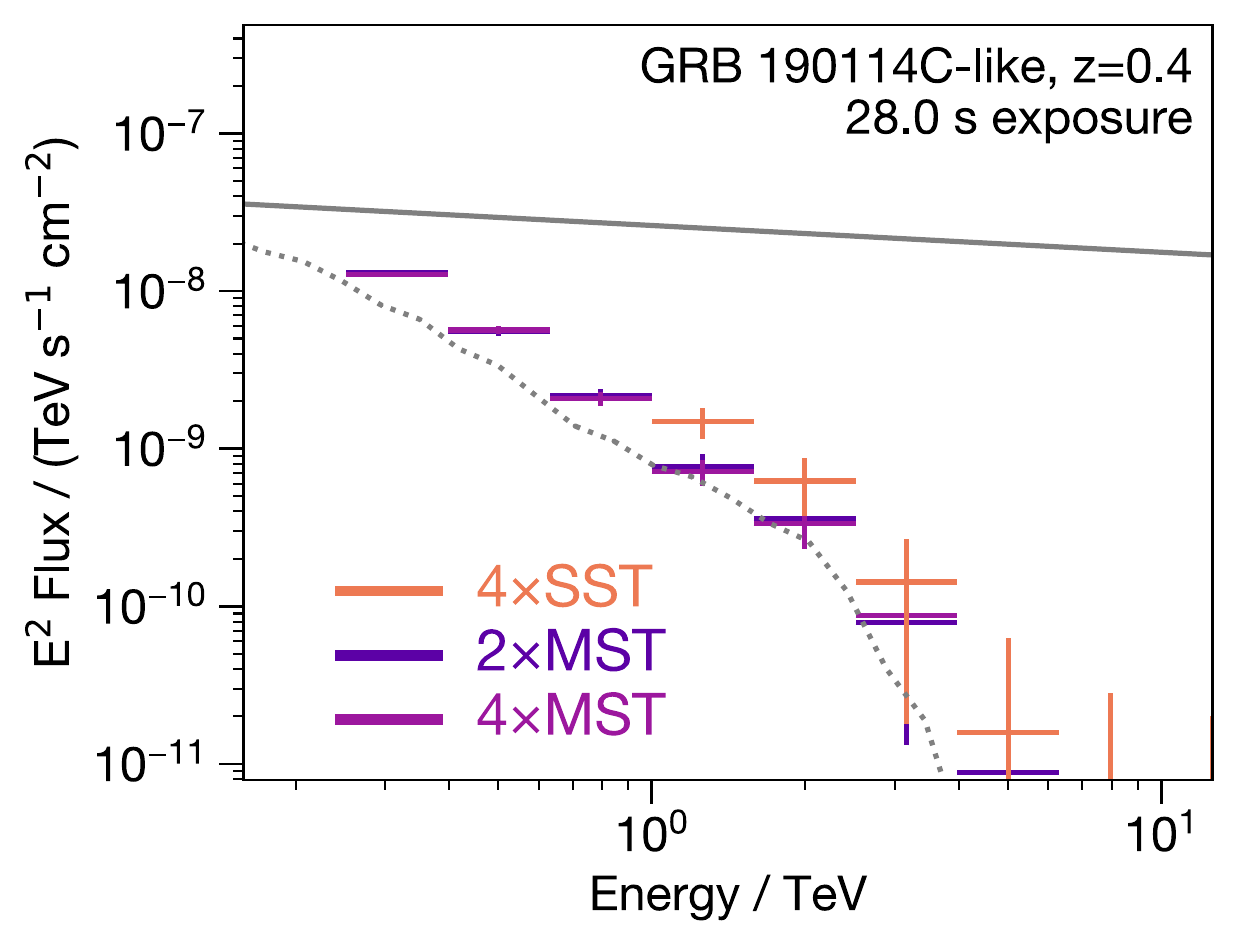}
          \caption{Reconstructed flux with EBL absorption for a GRB using counts estimation as described in \autoref{lightcurve} (0\,m altitude, 277\,m baselines).
          Intrinsic source flux (grey, solid) and source flux with EBL absorption (grey, dotted) are shown.
          The flux overestimation is due to energy dispersion and the very steep source spectrum.
          As such the lowest energy bins are not displayed.}
          \label{spectrum}
    \end{center}
\end{figure}

\begin{figure}[t]
    \begin{center}
          \centering
          \includegraphics[width=0.85\columnwidth]{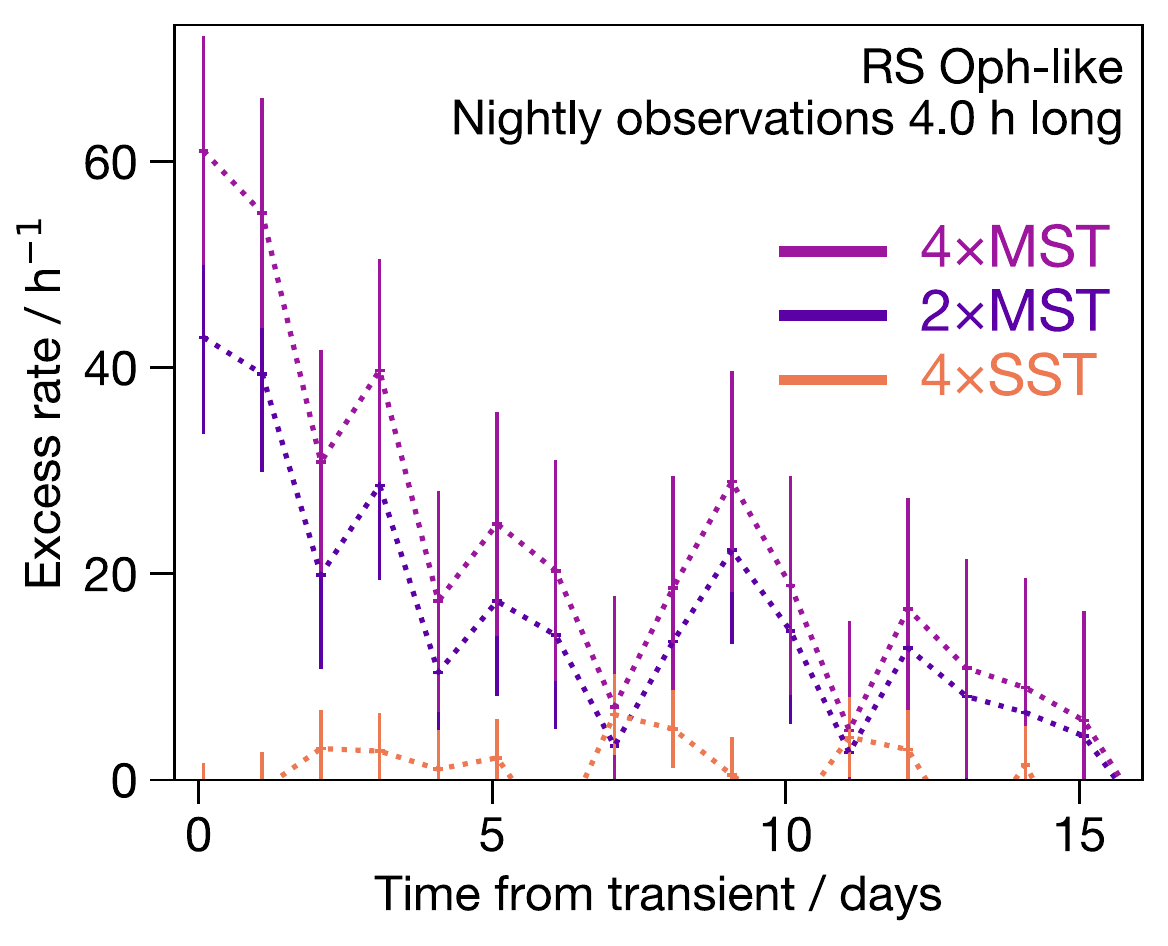}
          \caption{Estimated light curves for a flare akin to that from the recurrent nova RS Ophiuchi on the 8\textsuperscript{th} of August 2021 for arrays at 0\,m altitude with baselines of 277\,m.
          The mean background rates were 368/265/45 protons and electrons per hour for 4$\times$MST/2$\times$MST/4$\times$SST.
          The first 4\,h bin represents a 5.7$\sigma$ detection with four MSTs.}
          \label{novalightcurve}
    \end{center}
\end{figure}

The primary impact of site altitude on performance in this study was in the lower energy thresholds achievable at 1000\,m.
When observing sources with soft spectra or at high redshifts (with significant extragalactic background light (EBL) absorption) this is an important consideration.
For instance, the simulated lightcurves for a nova eruption (see \autoref{novalightcurve}) and GRB (see \autoref{lightcurve}) were almost identical for arrays of two MSTs at 0\,m altitude and four MSTs at 1000\,m altitude (both with 277\,m baselines).
The energy threshold improvements would need to be weighed against the extra costs associated with building and maintaining a high-altitude site.
The improved angular resolution available with three or four telescopes at large (277\,m) baselines (see \autoref{baseline}) is of note, as resolving the morphology of extended sources can be used to study their nature.
Physical space constraints at any given location may limit the ability to implement such a wide array, especially for high-altitude (possibly mountainous) regions.

One of the key objectives of a worldwide IACT network would no doubt be to follow up on GRBs to learn about their jet composition, central engines, and their mechanisms of radiation and energy dissipation \citep{Zhang2018}.
Figures \ref{lightcurve} and \ref{spectrum} and show estimated lightcurves and reconstructed spectra for an event akin to GRB 190114C (the first detected TeV GRB) for some example arrays and assumptions.
The on- and off-region counts were estimated using the gamma-ray and proton effective areas post-cuts derived from MC simulations, and the intrinsic flux of the source as measured by MAGIC in \citet{Veres2019}.
A cosmic-ray electron background was added, derived from the spectrum in \citet{HESS2017} convolved with the gamma-ray effective area and $\theta^2$ cuts.
To model the GRB flux decay over time, a relationship of $F(t) \propto  t^{-1.2}$ was assumed.
The EBL absorption model provided by \citet{Inoue2013} was implemented based on the GRB's redshift of $z\approx 0.4$, and additional random variation in counts was applied assuming Poissonian distributions.
Lastly, the RF models of energy reconstruction were used to disperse the counts between energy bins.
The suitability of MSTs for transient observations is clear, providing a significant advantage due to their lower energy threshold.
\autoref{shortgrblightcurve} additionally shows similar results for the observation of a short-duration GRB.
Creating densely sampled lightcurves of GRBs and tracking the change in their spectral energy distributions up to very high energies would help disambiguate the processes under which they are generated.

The recent detection of $>$100\,GeV gamma rays by H.E.S.S. and MAGIC from the recurrent nova RS Ophiuchi \citep{HESS2022, MAGIC2022} confirms novae as a new VHE gamma-ray source class and will undoubtedly spur further research into their properties.
\autoref{novalightcurve} shows estimated lightcurves of this flaring episode for some example arrays.
\emph{Fermi}-LAT daily light curves \citep{fermi2021} were fit with a spectral index of 1.9 \citep{cheung2021} to obtain flux normalisation and exponential temporal decay constants.
A spectral break was then applied at 13\,GeV, near the upper energy limit of \emph{Fermi}-LAT, with a spectral index of 4 used above the spectral break based on observations from H.E.S.S. \citep{wagner2021_2}.
These lightcurves show the suitability of an MST-based telescope array to successfully observe the VHE gamma-ray flux of such a source and to track its decay (and potentially its rise).
A number of novae emitting GeV gamma-rays have been detected by \emph{Fermi}-LAT \citep{Franckowiak2018}, and multiple different emission models have been suggested that are consistent with the available data.
Recent years have had 10 to 15 Galactic novae detected per annum \citep{mukai2021} (with the brightest seen by \emph{Fermi}-LAT) and roughly as many again from extragalactic sources.
The additional observation time and follow-up capability granted by a worldwide IACT network would extend spectral measurements up to higher energies, provide denser time sampling than currently available, and help determine what emission model most accurately describes these events.

We present this study as a proof-of-concept comparison of some major factors that influence the performance of a small gamma-ray telescope array.
As such, performance was evaluated for a single set of favourable conditions. 
Degradation in performance due to poor weather, brighter night sky background, or larger zenith angles were not investigated.
Conversely, improved performance could be expected with further optimisation of event reconstruction and the use of a stereo trigger.
The telescopes in this study were simulated to trigger event recording independently of each other, based on a number of pixels exceeding a certain trigger threshold within a time window.
With higher trigger thresholds, accidental triggers can be decreased at the cost of fewer recorded low-energy showers.
IACT arrays can be triggered at a much lower threshold if a stereo trigger is used, requiring coincident triggers from multiple individual telescopes.
This can also include topological information about the triggering pixels to further decrease accidental triggers \citep{LopezCoto2016}.
Implementing such a stereo trigger has been shown to successfully reduce the energy threshold achievable by an array of IACTs \citep{tridon2010}.
Sensitivity cut optimisation in this study was also restrained by the limited proton shower simulations available.
This was due to the extensive CPU time required to generate enough TeV showers that survived tight $\theta^{2}$ and gamma score cuts, as well as a sufficient amount of surviving low-energy showers.

\section{CONCLUSION}
In this study we investigated the performance attainable by a small array of Imaging Air Cherenkov Telescopes at an Australian site, and considered the viability of such a site for the observation of very-high-energy transients and variable sources.
The main results of this study are as follows:
\begin{itemize}
  \item For a given telescope design, whether it be MST-class or SST-class, there are substantial performance improvements with an increased number of telescopes ($\sim$2.5$\times$ better sensitivity for each doubling from 1 to 2 to 4 telescopes).
  \item Wider baseline distances up to 277\,m provide up to two-fold improvements in angular resolution, especially with 3 or 4 telescopes reaching $\sim$0.05\textdegree.
  \item The significantly lower energy threshold of an MST-class design ($\sim$300\,GeV) compared to that of an SST-class ($\sim$1.2\,TeV) makes them much more appropriate for observations of transients.
  \item Above the SST energy threshold, two SSTs have similar sensitivity to one MST, and four SSTs have similar sensitivity to two MSTs.
  \item Site altitude has a minimal impact on performance up to the heights available in Australia of $\sim$1000m; there is a small improvement in energy threshold, which is a significant consideration for soft-spectrum and high-redshift sources.
\end{itemize}

The lower energy threshold of MST-class telescope arrays makes them a more appropriate design for detecting transients at short timescales, such as the estimated 5.7$\sigma$ detection of an RS Ophiuchi-like nova eruption from a 4-hour observation with four MSTs  (see \autoref{novalightcurve}).
The fact that an appropriate number of SST-class telescopes can achieve equal or better performance as MST-class arrays at TeV energies is worth noting due to the significant difference in size and cost.
The large improvements in angular resolution available with wider baselines make it a worthwhile design consideration.
All other factors being equal, a higher altitude site would be preferable for the observations of transients, however, the benefits are small enough that they may be outweighed by practical factors such as site accessibility and available infrastructure.

Establishing an Australian IACT site and contributing to a worldwide network of IACTs would complement CTA and extend its capabilities.
By filling in the gamma-ray visibility gap in the Southern Hemisphere sky, it would increase the opportunity for providing triggers to CTA for more sensitive observations, and sources detected by CTA could be tracked continuously as their visibility passes between telescope sites.

\begin{acknowledgements}
This work was conducted in the context of the CTA Consortium.
We gratefully acknowledge the work of the CTA simulation and telescope teams for the MST and SST configurations.
We also thank Konrad Bernlöhr and Gernot Maier for their feedback.
This work was supported with supercomputing resources provided by the Phoenix HPC service at the University of Adelaide.
S.L. gratefully acknowledges support through the provision of an Australian Government Research Training Program Scholarship.
This paper has gone through internal review by the CTA Consortium.
\end{acknowledgements}

\bibliographystyle{pasa-mnras}
\bibliography{main}

\begin{appendix}
\counterwithin{figure}{section}
\counterwithin{table}{section}

\section{Direction reconstruction cuts}

Presented below are the impact distance cuts applied when using a combination of geometric and Random Forest models for direction reconstruction.

\begin{table}[!htb]
\caption{Impact distance (metres) from the centre of a subset of telescopes within an array to the shower core above which RF direction reconstruction performs better on average than geometric direction reconstruction.}
\centering
\begin{tabular}{c|cccc}
\hline
 & \multicolumn{2}{c}{MST} & \multicolumn{2}{c}{SST} \\
Array & 0m & 1000m & 0m & 1000m \\ \hline
\begin{tabular}[c]{@{}c@{}}2 tels,\\ 80m baseline\end{tabular} & 0 & 110 & 60 & 0 \\ \hline
\begin{tabular}[c]{@{}c@{}}2 tels,\\ 139m baseline\end{tabular} & 130 & 120 & 120 & 90 \\ \hline
\begin{tabular}[c]{@{}c@{}}2 + central tels,\\ 139m baseline\end{tabular} & 150 & 150 & 120 & 100 \\ \hline
\begin{tabular}[c]{@{}c@{}}2 tels,\\ 277m baseline\end{tabular} & 170 & 310 & 110 & 120 \\ \hline
\begin{tabular}[c]{@{}c@{}}2 + central tels,\\ 277m baseline\end{tabular} & 390 & 370 & 170 & 200 \\ \hline
\begin{tabular}[c]{@{}c@{}}3 or 4 tels,\\ 139m baseline\end{tabular} & 310 & 290 & 150 & 150 \\ \hline
\begin{tabular}[c]{@{}c@{}}3 or 4 tels,\\ 277m baseline\end{tabular} & 420 & 410 & 860 & 440 \\ \hline
\end{tabular}
\label{rfcuts}
\end{table}

\section{Sensitivity vs altitude and baseline}

Presented here are sensitivity curves for specific setups, showing the effects of altitude and baseline distance.
These follow the method described in \autoref{sensitivity}.
\\
\\
\begin{figure}[!htb]
    \begin{center}
          \centering
          \includegraphics[width=0.95\columnwidth]{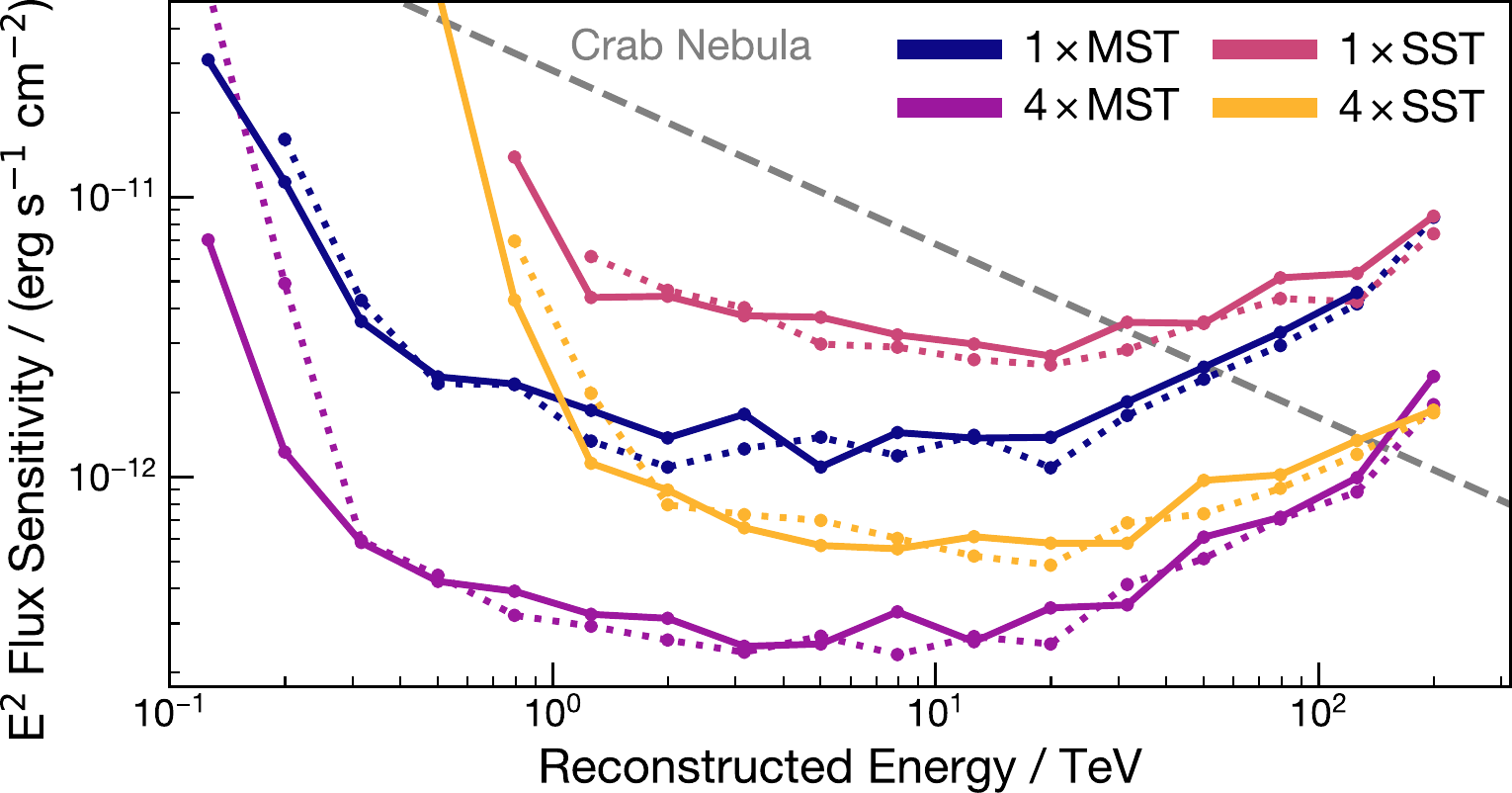}
          \caption{Sensitivity for 0\,m (dotted) and 1000\,m (solid) altitude arrays showing the improvement at low energies for the 1000\,m altitude arrays.
          4-telescope arrays had a 139\,m baseline.}
          \label{sens_alt}
    \end{center}
\end{figure}

\begin{figure}[!htb]
    \begin{center}
          \centering
          \includegraphics[width=0.95\columnwidth]{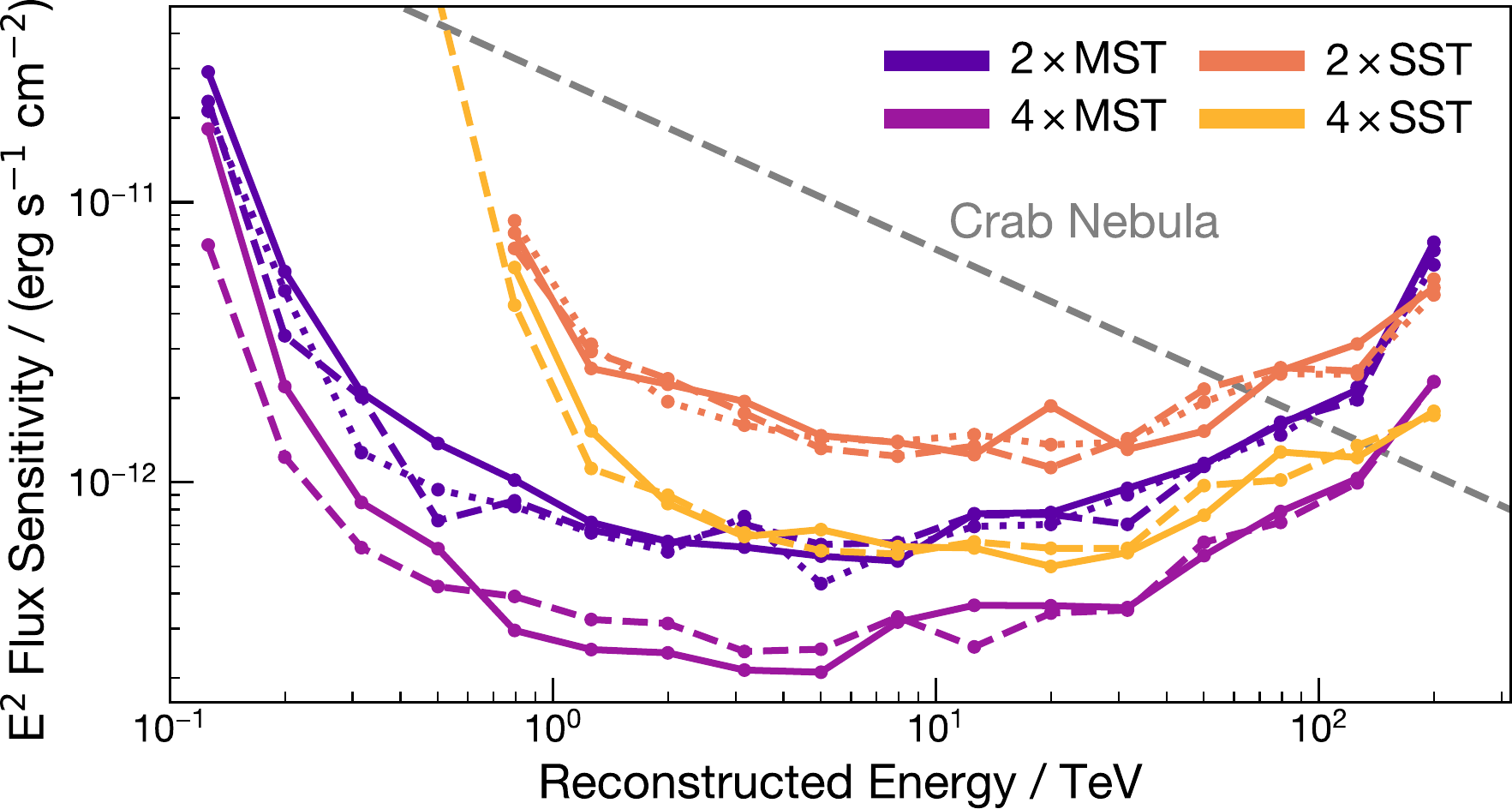}
          \caption{Sensitivity for 1000\,m altitude arrays with baselines of 80\,m (dotted), 139\,m (dashed), and 277\,m (solid) showing minimal differences in sensitivity performance due to baseline distance.}
          \label{sens_baseline}
    \end{center}
\end{figure}

\section{Short GRB lightcurves}

Presented here are simulated lightcurves for a GRB~160821B-like ``short GRB'' event.
While the flux quickly decays, this shows the suitability of a small MST array for detecting and monitoring such events.

\begin{figure}[!htb]
    \begin{center}
          \centering
          \includegraphics[width=0.85\columnwidth]{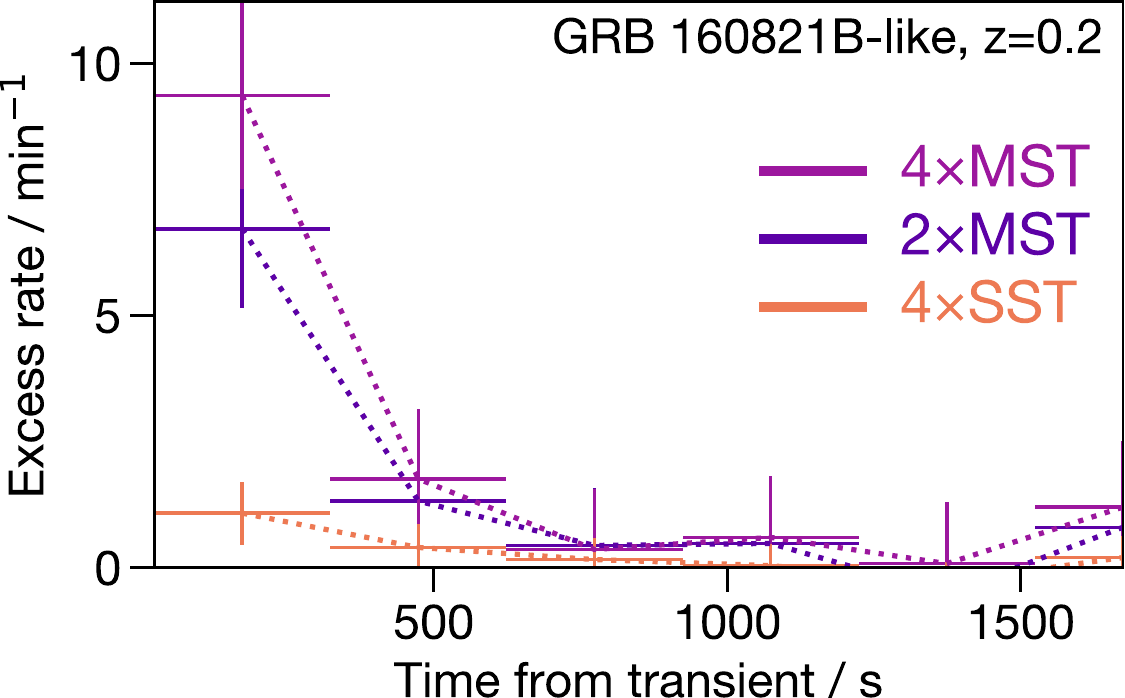}
          \caption{Simulated light curves for a GRB 160821B-like event for 0\,m altitude, 277\,m baseline arrays.
          Intrinsic source flux was based on the model in \citet{MAGICCollaboration2020} and scaled to match the flux seen by MAGIC, with temporal flux decay following $F(t) \propto  t^{-0.8}$.
          The mean background rates per bin were 6/4/1 protons and electrons per minute for 4$\times$MST/2$\times$MST/4$\times$SST.}
          \label{shortgrblightcurve}
    \end{center}
\end{figure}
\end{appendix}

\end{document}